\documentclass[12pt]{ronbun}

\usepackage{color,amsmath,amssymb,graphicx}

\usepackage{cite}
\usepackage{bm}
\usepackage{dcolumn}

\newcommand{\s}{\scriptscriptstyle}

\newcommand{\nn}{\nonumber}
\newcommand{\e}{{\rm e}}

\newcommand{\del}{\delta}

\newcommand{\ba}{\begin{eqnarray*}}
\newcommand{\ea}{\end{eqnarray*}}

\newcommand{\KUCPlogo}{\hbox{\lower 1.4ex\hbox{
\Huge\boldmath $\cal K$}
\kern -1.15em {\sffamily \bfseries\large\ UCP}}
\kern -4.5em \raise 0.2em\hbox{\lower 1.4ex\hbox{\color{cyan}
\Huge\boldmath $\cal K$}
\kern -1.15em {\color{magenta}\sffamily \bfseries\large\ UCP}
\put(-20,-7){\tiny\it preprint}
}}

\numberwithin{equation}{section}

\begin{document}

\begin{flushright}

\parbox{3.2cm}{
{KUCP-0214 \hfill \\
{\tt hep-th/0207190}}\\
\date }
\end{flushright}

\vspace*{0.5cm}

\begin{center}
 \Large\bf Giant Graviton and Quantum Stability \\
in Matrix Model on PP-wave Background
\end{center}

\vspace*{1.0cm}

\centerline{\large Katsuyuki Sugiyama$^{\ast}$ 
and Kentaroh Yoshida$^{\dagger}$}

\begin{center}
$^{\ast}$\emph{Department of Fundamental Sciences, \\
Faculty of Integrated Human Studies, \\
Kyoto University, Kyoto, 606-8501, Japan.} \\
{\tt E-mail:~sugiyama@phys.h.kyoto-u.ac.jp}\\
\vspace{0.2cm}
$^{\dagger}$\emph{Graduate School of Human and Environmental Studies,
\\ Kyoto University, Kyoto 606-8501, Japan.} \\
{\tt E-mail:~yoshida@phys.h.kyoto-u.ac.jp}
\end{center}

\vspace*{1.2cm}

\centerline{\bf Abstract}

We study classical solutions in 
Berenstein-Maldacena-Nastase (BMN) matrix model. 
A supersymmetric (1/2 BPS) fuzzy sphere is one of the classical solutions and 
corresponds to a giant graviton. We also consider 
other classical solutions, such as 
non-supersymmetric fuzzy sphere and harmonic oscillating gravitons. 
Some properties of oscillating gravitons are discussed. 
In particular, oscillating gravitons turn into 
usual supergravitons in the limit $\mu \rightarrow 0$. Moreover, 
we calculate the one-loop effective action  
around the supersymmetric fuzzy sphere by the 
use of the background field method and show 
the quantum stability of the giant graviton. Also, the instability of
the non-supersymmetric fuzzy sphere is proven.  

\vspace*{1.5cm}

\thispagestyle{empty}
\setcounter{page}{0}

\newpage 

\section{Introduction}

Recently, pp-wave backgrounds are so focused and intensively 
studied.  A maximally supersymmetric pp-wave background 
is a classical solution of the eleven-dimensional 
supergravity \cite{KG} and it is considered as one of the candidates for
supersymmetric backgrounds of M-theory \cite{OP}. 
The metric for this solution found by Kowalski-Glikman \cite{KG} 
(often called KG solution) is described by 
\begin{eqnarray}
ds^2 &=& - 2 dx^+ dx^- + G_{++}
(dx^+)^2 + (dx^{\mu})^2\, , \nn \\
& & G_{++} \equiv - \left[\left(\frac{\mu}{3}\right)^2 
 (x_1^2 + x_2^2 + x_3^2) + \left(\frac{\mu}{6}\right)^2 (x_4^2 + \cdots + 
 x_9^2)\right]\, , \nn  
\end{eqnarray}
and the constant 4-form flux for $+$, 1, 2 and 3 directions,    
\begin{equation}
 F_{+123} = \mu\, ,\quad (\mu \neq 0) \nn 
\end{equation}
is equipped. There are many other pp-wave backgrounds. 
In particular, after the maximally supersymmetric 
type IIB pp-wave solution has been 
found \cite{OP2}, it was shown that the type IIB string theory 
on the pp-wave is exactly-solvable in the Green-Schwarz (GS) 
formulation with the 
light-cone gauge fixing \cite{M,MT,RT}. The action of the string 
on the pp-wave acquires mass terms for bosons and fermions, 
but it is still free theory and  exactly-solvable. 
It should be noted that pp-wave backgrounds 
are curved and hence are interesting objects 
as clues to study string theories on the curved backgrounds, which are 
difficult subjects. 
The pp-wave background used in the type IIB analysis  
can be also obtained from the five-dimensional anti-de Sitter space $AdS_5$ 
through Penrose limit \cite{Penrose,Guven}. 
From this fact, the $AdS$/CFT correspondence has
been investigated \cite{Malda}. It is possible 
to study the $AdS$/CFT correspondence at the stringy level without 
reducing the analysis to the supergravity level in the region where 
the $AdS$ geometry can be approximated by the pp-wave geometry. 
It might be possibly expected 
that some new features of string theories could be understood due to
this advantage. 

It is well-known that the 
matrix theory approach to the M-theory seems very successful \cite{BFSS}.
Motivated by this, we can also study 
the Matrix theory \cite{Malda,DSR} or the supermembrane 
\cite{SY,SY2} on the pp-wave. 
The matrix model on the eleven-dimensional maximally supersymmetric 
pp-wave background has been proposed by D.~Berenstein, 
J.~Maldacena and H.~Nastase \cite{Malda}, which is 
often referred as the BMN matrix model. This model 
has mass terms and Myers terms (bosonic 3-point coupling terms) 
and contains more interesting physics than the flat case.  
The pp-wave background is curved and seems 
more complicated than the flat space but there are some
advantages. For an example, let us consider the supermembrane on the pp-wave. 
It is a well-known problem 
that a single supermembrane in the flat space is unstable
\cite{unstable}.  
However, possibly surprisingly, the quantum supermembrane on the pp-wave 
might be stable since  
flat directions of the quartic potential are 
completely removed due to the presence of mass terms.
The continuous spectrum of the supermembrane in the flat space 
is a troubling feature. 
However the spectrum of the BMN matrix model might be expected to be 
discrete and lead to an isolated set of the classical supersymmetric
vacua. This is convenient because the structure of the ground states 
is governed by the semi-classical approximation and it is not needed to
solve the full quantum mechanical problem of the ground state wave
function as in the flat space, which is too difficult.     
Thus, it can be expected in the BMN matrix model that some 
difficulties in the flat case are removed.

The BMN matrix model can be also derived from 
the supermembrane theory on the pp-wave through the matrix
regularization \cite{DSR}. We have discussed 
that the supermembrane theory on the maximally supersymmetric 
pp-wave is closely related to the BMN matrix model in the same way as 
the flat space in our previous works \cite{SY,SY2}. 
We have calculated the superalgebra in the supermembrane 
theory on the pp-wave background 
by using the standard Poisson-Dirac bracket procedure. 
The result agrees with the superalgebra in the BMN matrix model 
and confirms the correspondence of the superalgebra 
between the supermembrane and the matrix model on the
pp-wave. (The correspondence in the flat case has been discussed 
in \cite{BSS}.) We have also obtained the central charges (brane charges)
of the superalgebra, some of
which exist only in the pp-wave case.   
Moreover, we have investigated BPS conditions of the supermembrane on
the pp-wave and, for an example,  
constructed a 1/4 BPS solution whose charge equals to 
the angular momentum. It implies that this solution should be  
a rotating BPS configuration. BPS multiplets in the BMN matrix model are 
also intensively discussed in \cite{BPS}.

In this paper we consider classical solutions in the BMN matrix model. 
A supersymmetric (1/2 BPS) fuzzy sphere (giant graviton), 
a non-supersymmetric fuzzy sphere and 
oscillating gravitons are considered. 
The supergraviton in the flat space corresponds to a collection of 
harmonic oscillators on the pp-wave that is 
non-supersymmetric and considered as a non-BPS object. 
This result comes from the presence of bosonic mass terms. 
However, it turns into the supergraviton in the flat 
limit $\mu \rightarrow 0$.  Moreover, 
we calculate the one-loop effective action around the 
supersymmetric fuzzy sphere 
by using the background field method 
and prove the quantum stability of the giant graviton. 
The one-loop corrections are certainly canceled as expected from the 
requirement of the supersymmetry. 
However, it should be noted that its cancellation is non-trivial 
since the quantum supersymmetric vacua is non-trivial in the light-cone
formulation. It depends on the fact that 
the supercharges do not commute with the
Hamiltonian. In particular, the zero-point energy depends on the 
definition of the vacua. The one-loop contribution in the BMN matrix
model is essentially the zero-point energy contribution 
induced by the pp-wave background, and hence 
the cancellation of one-loop contributions would be also non-trivial.     
Also, we show the instability of the non-supersymmetric 
fuzzy sphere background in order to compare it with 
the supersymmetric background. 

This paper is organized as follows. 
Section 2 is devoted to the setup in this paper and  
we provide a brief review of the BMN matrix model. 
In section 3 we consider various classical solutions 
on the pp-wave. Besides a supersymmetric fuzzy sphere, 
we present a non-supersymmetric fuzzy 
sphere and harmonic oscillating gravitons 
corresponding to supergravitons in the flat space.
In section 4, we introduce the background field method in 
the BMN matrix model. The quadratic action is available to calculate  
the one-loop effective action. In section 5, we calculate the one-loop 
effective action around the supersymmetric fuzzy
sphere and prove the quantum stability of the giant graviton. 
We also discuss the instability of non-supersymmetric fuzzy sphere background. 
Section 6 is devoted to conclusions and discussions. 

\section{Berenstein-Maldacena-Nastase (BMN) Matrix Model}

In this section we briefly review the matrix theory on 
the eleven-dimensional maximally supersymmetric pp-wave 
background (BMN matrix model). 
The action of the BMN matrix model $\mathcal{S}$ is given by 
\begin{eqnarray}
\label{BMN}
\mathcal{S} &=& \mathcal{S}_{\s\rm flat} 
+ \mathcal{S}_{\mu}\,, \label{before} \\ 
 \mathcal{S}_{\s\rm flat} &=& \int\!dt \,{\rm Tr}\Bigg[
\frac{1}{2R}D_0X^rD_0X^r + \frac{R}{4}([X^r,\,X^s])^2 +
\Psi^{\s T}D_0\Psi + iR \Psi^{\s T}\gamma_r[\Psi,\,X^r]
\Bigg]\, ,\nn  \\ 
\label{BMN-1}
\mathcal{S}_{\mu} &=& \int\!dt \,{\rm Tr}\Bigg[
- \frac{1}{2R}\left(\frac{\mu}{3}\right)^2 X_{\s I}^2 - \frac{1}{2R}\left(
\frac{\mu}{6}\right)^2 X_{\s I'}^2 
- \frac{\mu}{3}i \epsilon_{\s IJK}X^{\s I}X^{\s J}X^{\s K}
- \frac{\mu}{4}\Psi^{\s T}\gamma_{\s 123}\Psi
\Bigg]\,. \nn 
\label{BMN-2}
\end{eqnarray}
where the covariant derivative is defined by 
\begin{eqnarray}
 D_0 X^r &\equiv& \partial_{t}X^r - i[A,\,X^r] \;\equiv\; 
\dot{X} -i [A,\,X^r]\,. 
\nn
\end{eqnarray}
When we shall rescale the gauge field $A$, parameters $t$ and $\mu$
as 
\[
 t \rightarrow \frac{1}{R}\,t \,,\quad 
A \rightarrow R A\,, \quad \mu \rightarrow R\,\mu \,,
\]
the actions $\mathcal{S}_{\s\rm flat}$ and $\mathcal{S}_{\mu}$ can 
be rewritten as 
\begin{eqnarray}
\label{act0}
\mathcal{S}_{\s\rm flat} &=& \int\!dt\, 
{\rm Tr}\left[\,
\frac{1}{2}D_0X^rD_0X^r + \frac{1}{4}([X^r,\,X^s])^2 + \Psi^{\s T}D_0\Psi
+ i\Psi^{\s T}\gamma_r[\Psi,\,X^r]
\right]\, , \\ 
\mathcal{S}_{\mu} &=& \int\!dt \,{\rm Tr}\Bigg[
- \frac{1}{2}\left(\frac{\mu}{3}\right)^2 X_{\s I}^2 - \frac{1}{2}\left(
\frac{\mu}{6}\right)^2 X_{\s I'}^2 
- \frac{\mu}{3}i \epsilon_{\s IJK}X^{\s I}X^{\s J}X^{\s K}
- \frac{\mu}{4}\Psi^{\s T}\gamma_{\s 123}\Psi
\Bigg]\,. 
\label{act1}
\end{eqnarray}
This theory has 16 dynamical supersymmetries, 
\begin{eqnarray}
 \del_{\epsilon}X^r &=& 2\psi^{\s T}\gamma^{r}\epsilon(t)\,,\quad 
 \del_{\epsilon}A \;=\; 2\psi^{\s T}\epsilon(t)\,, \nn \\
 \del_{\epsilon}\psi &=& \left[D_0X^r\gamma_r + \frac{i}{2}[X^r,\,X^s]
\gamma_{rs} + \frac{\mu}{3}\sum_{{\s I}=1}^3X^{\s I}\gamma_{\s I}
\gamma_{\s 123} - \frac{\mu}{6}\sum_{{\s I'}=4}^9X^{\s I'}\gamma_{\s I'}
\gamma_{\s 123}
\right]\epsilon(t)\,,  \label{SUSY} \\ 
\epsilon(t) &=& \e^{- \frac{\mu}{12}\gamma_{\s 123}t}\epsilon_0\,,\quad 
\epsilon_0:~\mbox{constant spinor}\,, \nn 
\end{eqnarray}
and 16 kinematical supersymmetries, 
\begin{eqnarray}
\del_{\eta}X^r &=& \del_{\eta}A \;=\; 0\,, \nn \\
\del_{\eta}\psi &=& \eta(t)\,, \label{SUSY2} \\
\eta(t) &=& \e^{\frac{\mu}{4}\gamma_{\s 123}t}\eta_0\,,\quad \eta_0:~\mbox{constant spinor}\,. \nn 
\end{eqnarray}
Note that the bosons and fermions in the action $\mathcal{S}$ have 
different masses. Three of bosons have mass $\mu/3$ and the remains 
have mass $\mu/6$. On the other hand, all the fermions have mass
$\mu/4$. This result depends on the fact that 
the supercharge do not commute with the Hamiltonian. It might be 
considered that such a situation comes 
from the light-cone gauge fixing \footnote{The covariant approach is
also discussed in Ref.\,\cite{HKS}.}.

\section{Classical Solutions  of BMN Matrix Model}

In this section we shall consider classical solutions 
by solving classical equations of motion under a certain ansatz.  
In particular, a fuzzy sphere solution preserving a half of 
supersymmetries (i.e., 1/2 BPS), a non-supersymmetric fuzzy sphere
solution, and oscillating gravitons are constructed here 
\footnote{Other several interesting solutions have been  
reported in \cite{bak}, though we will not consider them in this paper.}. 

Let us set fermionic degrees of freedom to zero since 
we are restricted ourselves to bosonic backgrounds. 
Taking the variations of $X$'s in the action $\mathcal{S}$,  
we obtain classical equations of motion, 
\begin{eqnarray}
\label{eq0}
\ddot{X}^{\s I} &=& -[[X^{\s I},\,X^r],\,X^r] 
- \left(\frac{\mu}{3}\right)^2X^{\s I} 
- i\mu \epsilon_{\s IJK}X^{\s J}X^{\s K}\,, \quad (I = 1,\,2,\,3), 
\\
\ddot{X}^{\s I'} &=& -[[X^{\s I'},\,X^r],\,X^r] 
- \left(\frac{\mu}{6}\right)^2 X^{\s I'}\,,\quad (I' = 4,\,\ldots,\,9). 
\label{eq1}
\end{eqnarray}
In order to show the energy of classical solutions, 
we utilize the bosonic potential $V_B$ in the BMN matrix model defined by 
\begin{eqnarray}
 V_{\rm\s B}(X) &=& 
{\rm Tr}\left[
- \frac{1}{4}([X^r,\,X^s])^2  + \frac{1}{2}\left(\frac{\mu}{3}\right)^2
X_{\s I}^2 + \frac{1}{2}\left(\frac{\mu}{6}\right)^2 X_{\s I'}^2
+ \frac{\mu}{3}
i\epsilon_{\s IJK}X^{\s I}X^{\s J}X^{\s K}\right]\,, \nn \\
&=& \frac{1}{2}\,{\rm Tr}\Bigg[\left(\frac{\mu}{3}X^{\s I} + i\epsilon_{\s IJK} X^{\s J}X^{\s K} \right)^2 \nn \\
&& + \frac{1}{2}(i[X^{\s I'},\,
X^{\s J'}])^2 + (i[X^{\s I'},\,X^{\s I}])^2 + \left(\frac{\mu}{6}\right)^2
(X^{\s I'})^2 \Bigg]\,.
\end{eqnarray}
The supersymmetry transformations (\ref{SUSY}) and 
(\ref{SUSY2}) are also available to check whether classical solutions are 
supersymmetric or not.

\subsection{Giant Graviton and Fuzzy Sphere Solution}

We would like to obtain static fuzzy sphere solutions of 
equations of motion (\ref{eq0}) and (\ref{eq1}). To begin,  
we shall suppose the following ansatz, 
\begin{equation}
 X^{\s I} \;=\; \alpha J^{\s I}\,,\quad X^{\s I'} \;=\; 0\,, 
\label{an}
\end{equation}
where $\alpha$ is an arbitrary constant and $J^{\s I}$'s are $SU(2)$
generators and satisfy an $SU(2)$ Lie algebra, 
\[
 [J^{\s I},\,J^{\s J}] \;=\; i\epsilon_{\s IJK}J^{\s K}\,.
\]
In this case, Eq.\,(\ref{eq1}) is satisfied automatically.  
Inserting the ansatz (\ref{an}) into Eq.(\ref{eq0}), we obtain 
an algebraic equation, 
\[
 \alpha\left(\alpha - \frac{\mu}{3}\right)\left(\alpha -
 \frac{\mu}{6}\right) \;=\; 0\,.
\]
The case $\alpha = 0$ denotes a trivial classical supersymmetric vacuum, 
$X^r = 0$. The value $\alpha = \mu/3$ corresponds to 1/2 BPS fuzzy sphere
solution as pointed out in Ref.\,\cite{Malda}.  
This solution can be interpreted 
as a giant graviton, which is the expanded D0-branes due to the 
Myers effect in the presence of the constant 4-form flux \cite{Myers}. 
In general a giant graviton has finite energy 
but the energy of this solution is zero   
as we can easily see by inserting the solution into the bosonic
potential $V_{\rm\s B}$. 
This solution is labelled by all possible ways of dividing 
an $N$-dimensional representation of $SU(2)$ into irreducible
representations. This number is equal to that of partitions of $N$. 
This is also the number of multiple graviton states with zero energy. 
Each irreducible representation corresponds to a single graviton and
hence several gravitons exist in the 
supersymmetric vacua. Thus the structure of the classical supersymmetric 
vacua is non-trivial and interesting \cite{DSR}. 

Moreover, another fuzzy sphere solution $\alpha = \mu/6$ 
is allowed as a classical
solution of the equations of motion. 
This solution is non-supersymmetric and has finite energy, 
\[
 E \;=\; \frac{\mu^4}{2592}{\rm
 Tr}\left[(J^{\s I})^2\right]\,.
\] 
The above expression of the energy is described by the quadratic 
Casimir, and hence it depends on the representations of the 
generator $J^{\s I}$. The physical interpretation is 
unclear but it might be possibly interpreted as a kind of 
the extended objects due to the Myers effect.  
Finally, we should comment that 
this non-supersymmetric fuzzy sphere is unstable at the one-loop level, 
as we will see later. 

\subsection{Oscillating Gravitons on the PP-wave}

In the flat space it is understood how 
various M-theory objects, 
such as M2-brane (supermembrane), M5-brane and supergraviton 
can be constructed in the context of the matrix theory. 
In the pp-wave case several additional terms exist 
in the action and equations of motion are modified. 
Thus, the dynamics is drastically changed 
even at the classical level. We can guess  
some modifications for the dynamics of such M-theory objects  
\footnote{In fact, we have discussed 
supermembranes in our previous works 
\cite{SY,SY2}.}.  In this subsection we shall 
consider an object in the BMN matrix model corresponding to the 
supergraviton on the flat background. 
In the flat space the 
discrete light-cone quantized (DLCQ) M-theory 
should have a point-like state corresponding 
to a longitudinal graviton with $p^+ =N/R$ and arbitrary transverse
momenta $p^r$'s \footnote{In this subsection, 
we use the action (\ref{before}). 
} $(r=1,2,\cdots,9)$. This is the supergraviton. 
In the same way, we can consider 
an object on the pp-wave corresponding to the supergraviton and 
it seems an interesting object to study. We  will construct  
such solutions explicitly below. 

One simple family of solutions for these equations of motion (\ref{eq0})
and (\ref{eq1}) is the type with the diagonal form. 
In this case the term with commutators and Myers
term vanish, and equations of
motion are reduced to those of the 
harmonic oscillator. Then this type of solution is given by  
\begin{eqnarray}
X^{\s I} &=& 
\begin{pmatrix}
\, x^{\s I}_1\cos\left(\frac{\mu}{3}t \right)
+ \frac{3}{\mu}v^{\s I}_1 \sin\left(\frac{\mu}{3}t\right) 
 & &   \\
 & \ddots     &      \\
 &  & x^{\s I}_N \cos\left(\frac{\mu}{3}t\right)
+ \frac{3}{\mu}v^{\s I}_N \sin\left(\frac{\mu}{3}t \right)
\end{pmatrix}\,,
\end{eqnarray}
for $I=1,\,2,\,3$, and 
\begin{eqnarray}
X^{\s I'} &=& 
\begin{pmatrix}
\, x^{\s I'}_1 \cos\left(\frac{\mu}{6}t\right) 
+ \frac{6}{\mu}v^{\s I'}_1\sin\left(\frac{\mu}{6}t\right) &  &  \\
  &  \ddots     &      \\
   &  & x^{\s I'}_N \cos\left(\frac{\mu}{6}t\right)
+ \frac{6}{\mu}v^{\s I'}_N \sin\left(\frac{\mu}{6}t\right)
\end{pmatrix}\,,
\end{eqnarray}
for $I'=4,\,\ldots,\,9$. It can be easily shown that this harmonic 
oscillating solution turns into an $N$-supergraviton solution in the  
flat limit $\mu \rightarrow 0$. Thus this type of classical 
solution corresponds to the supergraviton in the flat space 
and in this sense we refer to this type of solution 
as the oscillating $N$-graviton.  One can also show that 
it is non-supersymmetric and has finite energy, 
though a supergraviton in the flat space 
is a supersymmetric (1/2 BPS) object. 
It would be possibly guessed that a BPS saturated supergraviton 
might be lifted up to the 
non-BPS state due to 
the effect induced by the pp-wave background, such as 
bosonic mass terms. 

Each graviton has respectively the longitudinal momentum $p_a^+$ 
and transverse one  
$p_a^{r}$ $(a=1,\,\ldots,\,N)$ defined by 
\begin{eqnarray}
 p_a^+ \;=\; \frac{1}{R}\,, \qquad p_a^{r} \;=\; v_a^r\frac{1}{R}\,,\quad 
(r=1,2,\cdots ,9).
\end{eqnarray} 
Its associated energy is expressed as 
\begin{eqnarray}
E_a &=& \frac{1}{2R}(v_a^r)^2  + \frac{1}{2R}\left(\frac{\mu}{3}\right)^2
\sum_{{\s I}=1}^3(x_a^{\s I})^2 + \frac{1}{2R}
\left(\frac{\mu}{6}\right)^2 \sum_{{\s I'}=4}^9(x_a^{\s I'})^2 \nn \\
&=& \frac{1}{2p^+_a}(p_a^r)^2 + \frac{1}{2}p^+_a 
\left(\frac{\mu}{3}\right)^2
\sum_{{\s I}=1}^3(x_a^{\s I})^2 + 
\frac{1}{2}p^+_a 
\left(\frac{\mu}{6}\right)^2 \sum_{{\s I'}=4}^9(x_a^{\s I'})^2\, .
\end{eqnarray}
The ``mass'' is $p^+_a$ and 
its ``spring constant'' is proportional to $\mu^2$ which arises due to 
the effect of the pp-wave background.  
In the $\mu\rightarrow 0$ limit, we can recover the result in the flat
space where the supergraviton solution is a non-relativistic free
particle with ``mass'' $p^+_a$.  

Also, a single classical graviton with the longitudinal momentum $p^+=N/R$ 
can be described by taking 
\[
 x^r_1 = \cdots = x^r_N\,, \quad v_1^r = \cdots = v_N^r\,,
\]
so that the harmonic oscillations of all matrix components are identical. 

Finally, we shall briefly comment on the quantum supergraviton. 
When the quantum supergraviton is clarified, we can know a number of
states in the supergravity multiplet. In the case of the flat space, 
the supergraviton is certainly 
a member of the supergravity multiplet 
but has some difficulties such as the ground states problem. 
Besides such problems, 
members of the supergravity multiplet on the pp-wave have 
different masses  
in contrast with the flat case. 
This situation arises due to the fact that 
supercharges do not commute with the Hamiltonian and rather 
act as raising or lowering operators. 
It is the reason why the bosons and fermions in the BMN matrix
model can have different masses but the energy 
shifts of the members in the multiplet are still 
constrained by the supersymmetry algebra. Thus, it might be not 
surprising that the supergraviton acquires mass at the quantum level.  
Of course, the massless quantum supergraviton should be recovered 
in the limit $\mu\rightarrow 0$.

\section{Background Field Method}

In this section we explain the background field method in the BMN
matrix model in order to study whether the fuzzy sphere background is 
stable or not.

To begin, let us decompose the $X^r$ and $\Psi$ 
into the backgrounds $B^r,\, F$ 
and fluctuations $Y^r,\, \psi$ as 
\begin{eqnarray}
 X^r &=& B^r + Y^r \,,\quad  \Psi \;=\; F + \psi\,,
\label{deco}
\end{eqnarray} 
where we take $F=0$ since 
the fermionic background is not considered in this paper. 
Next, we shall take the background field gauge using 
the gauge-fixing terms and Faddeev-Popov ghost terms as 
\begin{eqnarray}
\mathcal{S}_{\s\rm GF+FP} \;=\; 
- \frac{1}{2}\int\!dt \,{\rm Tr}\Bigg[
(D_0^{\rm bg}A)^2 
+ i \bar{C}\partial_{t}D_{0}C + \bar{C}[B^r,\,[X^r,\,C]]
\Bigg]\,,
\end{eqnarray}
where $D_0^{\rm bg}A$ is defined by 
\begin{eqnarray}
 D_0^{\rm bg}A &=& \partial_tA  + i[B^r,\,X^r]\,.
\end{eqnarray}
The background field gauge condition is 
$D_0^{\rm bg}A = 0$. 
The advantage of this gauge choice is that 
the second order actions with respect to the fluctuations 
are simplified. 

Hereafter, we shall consider in the Euclidean formulation by 
taking $t \rightarrow i\tau$ 
and $A \rightarrow -iA$. When we insert the decompositions 
of the fields (\ref{deco}) 
into Eq.\,(\ref{BMN}), 
the action $\mathcal{S}$ is expanded around the background. 
The resulting action can be written as 
\begin{eqnarray}
 \mathcal{S} &=& \mathcal{S}_0 + \mathcal{S}_2 
+ \mathcal{S}_3 + \mathcal{S}_4 \,,
\end{eqnarray}
\begin{eqnarray}
\mathcal{S}_0 &=& \int\!d\tau {\rm Tr}\Bigg[\,
\frac{1}{2}(\dot{B}^r)^2 - \frac{1}{4}([B^r,\,B^s])^2 + \frac{1}{2}\left[
\left(\frac{\mu}{3}\right)^2(B^{\s I})^2 + \left(\frac{\mu}{6}\right)^2 
(B^{\s I'})^2 
\right] + \frac{\mu}{3}i\epsilon_{\s IJK}B^{\s I}B^{\s J}B^{\s K}
\Bigg]\,, \nn \\
\mathcal{S}_2 &=& \int\!d\tau{\rm Tr}\Biggl[\,
\frac{1}{2}(\dot{Y}^r)^2 - 2i\dot{B}^r 
[A,\,Y^r] - \frac{1}{2}([B^r,Y^s])^2 
- [B^r,\,B^s][Y^r,\,Y^s]
+ \mu i\epsilon_{\s IJK}B^{\s I}Y^{\s J}Y^{\s K} 
\nn \\
&&  
+ \frac{1}{2}\left[\left(\frac{\mu}{3}\right)^2Y_{\s I}^2 
+ \left(\frac{\mu}{6}\right)^2Y_{\s I'}^2 \right] 
+ i\psi^{\s T}\dot{\psi}
- i\psi^{\s T}\gamma_r[\psi,\,B^r] 
+ \frac{\mu}{4}\psi^{\s T}\gamma_{\s 123}\psi  \nn \\ 
&& + \frac{1}{2}\dot{A}^2  - \frac{1}{2}([B^r,\,A])^2 
+ \frac{1}{2}\dot{\bar{C}}\dot{C} - \frac{1}{2} [B^r,\,\bar{C}][B^r,
\,C] \Biggr]
\, ,  \label{sec}
\\
\mathcal{S}_3 &=& \int\!d\tau {\rm Tr}\Bigg[
- i\dot{Y}^r[A,\,Y^r] - [A,\,B^r][A,\,Y^r] -
[B^r,\,Y^s][Y^r,\,Y^s] +  \psi^{\s T}[A,\,\psi] \nn \\
&& - i\psi^{\s T}\gamma_r[\psi,\, Y^r] 
+ \frac{\mu}{3}i\epsilon_{\s IJK}Y^{\s I}Y^{\s J}Y^{\s K} 
- \frac{i}{2}\dot{\bar{C}}[A,\,C] 
- \frac{1}{2}[B^r,\,\bar{C}][Y^r,\,C] 
\Bigg]\,, \nn \\
\mathcal{S}_4 &=& 
\int\!d\tau {\rm Tr}\left[ -\frac{1}{2}([A,\,Y^r])^2 - \frac{1}{4}
([Y^r,\,Y^s])^2 
\right]\,, \nn 
\end{eqnarray}
where $\mathcal{S}_i$ is the order $i$ part with respect to 
fields representing fluctuations.
The symbol ``dot'' ($\cdot$) is redefined by tau ($\tau$) derivative as 
$\dot{A} \equiv \partial_{\tau}A$. 
The additional terms proportional to $\mu$ 
cannot appear in the quartic terms. 
The background field $B^r$ 
is not specified here yet. The classical solutions discussed
in the previous section can be taken as the background $B^r$.

\section{Quantum Stability of Giant Graviton}

We shall investigate whether the supersymmetric fuzzy
sphere (giant graviton) background and fuzzy sphere solution with 
no supersymmetry are stable or not 
by calculating the one-loop effective actions  
around these fuzzy spheres solutions \footnote{In the case of 
the usual type IIB matrix 
model, the fuzzy sphere background has been discussed in \cite{Iso}. }.

\subsection{Stability of Giant Graviton}

In this subsection we consider the quantum stability of the giant 
graviton that is a supersymmetric (1/2 BPS) fuzzy sphere.  
In this case the background should be taken as 
\begin{eqnarray}
 B^{\s I} \;=\; \frac{\mu}{3}J^{\s I}\,,\quad B^{\s I'} \;=\; 0\,.
\end{eqnarray}
We shall restrict ourselves to 
the $N=2$ case (i.e., $2\times 2$ matrices) for simplicity, 
and expand the fluctuations and gauge fields as 
\begin{eqnarray}
 Y^r &=& \frac{1}{\sqrt{2}}Y_0^{r}
{\bf 1}_2 + \sqrt{2}Y_1^{r}J^1 + \sqrt{2}Y_2^{r}J^2 
+ \sqrt{2}Y_3^{r}J^3\,, \nn \\
 \psi &=& \frac{1}{\sqrt{2}}\psi_0
{\bf 1}_2 + \sqrt{2}\psi_1 J^1 + \sqrt{2}\psi_2 J^2 
+ \sqrt{2}\psi_3J^3\,, \nn \\
 A &=& \frac{1}{\sqrt{2}}A_0{\bf 1}_2 + \sqrt{2}A_1J^1 + \sqrt{2}A_2J^2 
+ \sqrt{2}A_3J^3\,, \nn 
\end{eqnarray}
where $J^{\s I} \equiv \sigma^{\s I}/2,\, (I = 1,\,2,\,3) $  and
$\sigma^{\s I}$'s are Pauli matrices. The orthonormal relations and
quadratic Casimir of the 
$SU(2)$ generators $J^{\s I}$'s are given by  
\[
 {\rm Tr}[J^{\s I}J^{\s J}] \;=\; \frac{1}{2}\del^{\s IJ}\,,\quad 
 {\rm Tr}\left[(J^{\s I})^2\right] \;=\; \frac{3}{2}\,.
\]
These are useful to calculate the effective action.

Inserting these expansions into the second order action (\ref{sec}), 
we obtain the resulting action described by 
\begin{eqnarray}
 \mathcal{S}_2 &=& \int\!d\tau\,\Biggl[
\frac{1}{2}(\dot{Y}_0^{\s I})^2 
+ \frac{1}{2}\left(\frac{\mu^2}{9}\right)(Y_0^{\s I})^2 
+ \frac{1}{2}\left((\dot{Y}_1^{\s I})^2 
+ (\dot{Y}_2^{\s I})^2 
+ (\dot{Y}_3^{\s I})^2 \right) \nn \\
&& 
- \frac{\mu^2}{9}
(Y_1^1Y_2^2 + Y_2^2Y_3^3 + Y_3^3Y_1^1 - Y_1^2Y_2^1 - Y_3^2Y_2^3 
- Y_1^3Y_3^1) 
\nn \\
&&
+ \frac{1}{2}\left(\frac{\mu^2}{3}\right)
\left((Y_1^{\s I})^2 + (Y_2^{\s I})^2 + (Y_3^{\s I})^2 \right) 
+ \frac{1}{2}(\dot{Y}_0^{\s I'})^2 
+ \frac{1}{2}\left(\frac{\mu^2}{36}\right) (Y_0^{\s I'})^2 
\nn \\
&&
+ \frac{1}{2}\left((\dot{Y}_1^{\s I'})^2 
+ (\dot{Y}_2^{\s I'})^2 
+ (\dot{Y}_3^{\s I'})^2\right) 
+ \frac{1}{2}\left(\frac{\mu^2}{4}
\right)\left((Y_1^{\s I'})^2 + (Y_2^{\s I'})^2 + (Y_3^{\s I'})^2 
\right) \nn \\ 
&& + \frac{1}{2}\left((\dot{A_0})^2 + 
(\dot{A}_1)^2 + (\dot{A}_2)^2 
+ (\dot{A}_3)^2 \right) 
+ \frac{1}{2}\left(\frac{2}{9}\mu^2\right)\left(
(A_1)^2 + (A_2)^2 + (A_3)^2 
\right) \nn \\
&& + \frac{1}{2}\left(\dot{\bar{C}}_0 \dot{C}_0 
+ \dot{\bar{C}}_1\dot{C}_1 + \dot{\bar{C}}_2\dot{C}_2 
+ \dot{\bar{C}}_3\dot{C}_3\right) + \frac{1}{2}\left(\frac{2}{9}\mu^2\right)
\left(\bar{C}_1C_1 + \bar{C}_2C_2 + \bar{C}_3C_3
\right) \nn \\ 
&& + i\left(\psi_0\dot{\psi}_0 + \psi_1\dot{\psi}_1 
+  \psi_2\dot{\psi}_2 + \psi_3\dot{\psi}_3 \right) 
+ \frac{\mu}{4}\left(\psi^{\s T}_0\gamma_{\s 123}\psi_0 
+ \psi^{\s T}_1\gamma_{\s 123}\psi_1 
+ \psi^{\s T}_2\gamma_{\s 123}\psi_2
+ \psi^{\s T}_3\gamma_{\s 123}\psi_3 \right) \nn \\
&& - \frac{\mu}{3}\left(
\psi_1^{\s T}\gamma_2\psi_3 
- \psi_1^{\s T}\gamma_3\psi_2 
+ \psi_2^{\s T}\gamma_3\psi_1 
- \psi_2^{\s T}\gamma_1\psi_3 
+ \psi_3^{\s T}\gamma_1\psi_2 
- \psi_3^{\s T}\gamma_2\psi_1 
\right)
\Biggr]\,. 
\label{main}
\end{eqnarray} 
The diagonal parts of the action (\ref{main}) can be 
integrated out, but we must diagonalize the non-diagonal 
nine components $Y_i^{\s I}$'s,\,$(i,I = 1,\,2,\,3)$ and fermions
$\psi_1,\,\psi_2$ and $\psi_3$ in order to integrate out them. 
The diagonalization of the $9 \times 9$ matrix 
can be easily done (see, Appendix A for the detailed calculations). 
The resulting action for these 9 components are evaluated as
\begin{eqnarray}
 \mathcal{S}_{Z} &=& \int\!d\tau\, 
\Biggl[
\frac{1}{2}\left(\dot{Z}^1\right)^2 + \frac{1}{2}\left(\frac{\mu^2}{9}
\right) (Z^1)^2 \nn \\
&& + \sum_{i=2}^4\left[\frac{1}{2}(\dot{Z}^i)^2 
+ \frac{1}{2}\left(\frac{2}{9}\mu^2\right) (Z^i)^2
\right] 
+ \sum_{i=5}^9 \left[\frac{1}{2}(\dot{Z}^i)^2 
+ \frac{1}{2}\left(\frac{4}{9}\mu^2\right) (Z^i)^2
\right] 
\Biggr]\,.
\end{eqnarray}
Next, let us diagonalize the fermion parts. 
This can be also done after simple calculations (see, Appendix B for
the technical details), and 
the result is expressed
\begin{eqnarray}
 S_{\varphi} &=& \int\!d\tau\, \Biggl[
 i\varphi_0^{\s T}\dot{\varphi}_0 
+ i\varphi_1^{\s T}\dot{\varphi}_1 
+ i\varphi_2^{\s T}\dot{\varphi}_2 
+ i\varphi_3^{\s T}\dot{\varphi}_3 \nn \\ 
&& + \frac{\mu}{4}\varphi_0^{\s T}\gamma_{\s 123}\varphi_0
+ \frac{7}{12}\mu\varphi_1^{\s T}\gamma_{\s 123}\varphi_1 
+ \frac{7}{12}\mu\varphi_2^{\s T}\gamma_{\s 123}\varphi_2
- \frac{5}{12}\mu\varphi_3^{\s T}\gamma_{\s 123}\varphi_3
\Biggr]\,,
\end{eqnarray} 
where $\varphi_i$'s are real spinors with 16 components. 

Finally, we can obtain the following contents of 
fields\footnote{We consider the case that $\mu$ is positive since we can
impose this condition without the loss of generality.}:
\vspace*{0.5cm}
\\
\underline{Bosonic Fluctuation $Y^r$}
\begin{itemize}
\item 4 massive bosons with mass $\displaystyle \frac{\mu}{3}$:\quad
      $Y_0^{\s I}$ and $Z^1$\,, 
\item 3 massive bosons with mass $\displaystyle \frac{\sqrt{2}}{3}\mu$:
\quad $Z^i\quad (i=2,\,3,\,4)$\,,
\item 5 massive bosons with mass $\displaystyle \frac{2}{3}\mu$:
\quad  $Z^i\quad (i=5,\,\ldots,\,9)$\,,
\item 6 massive bosons with mass $\displaystyle \frac{\mu}{6}$:
\quad $Y_0^{\s I'}$\,,
\item 18 massive bosons with mass $\displaystyle \frac{\mu}{2}$:
\quad $Y_{i}^{\s I'}\quad 
(i=1,\,2,\,3)$\,,
\end{itemize}
\underline{Gauge Field $A$}
\begin{itemize}
\item 1 massless boson:\quad $A_0$\,,
\item 3 massive bosons with mass $\displaystyle \frac{\sqrt{2}}{3}\mu$:
\quad $A_{i} \quad (i=1,\,2,\,3) $\,,
\end{itemize}
\underline{Ghost $C$}
\begin{itemize}
\item 1 massless complex ghost:\quad $C_0$\,,
\item 3 massive complex 
ghosts with mass $\displaystyle \frac{\sqrt{2}}{3}\mu$:
\quad $C_i \quad (i=1,\,2,\,3)$\,,
\end{itemize}
\underline{Fermion $\varphi$}
\begin{itemize}
\item 16 massive fermions with mass $\displaystyle \frac{\mu}{4}$:
\quad $\varphi_0$\,,
\item 32 massive fermions with mass $\displaystyle \frac{7}{12}\mu$:
\quad $\varphi_1$ and $\varphi_2$\,,
\item 16 massive fermions with mass $\displaystyle -\frac{5}{12}\mu$:
\quad $\varphi_3$\,.
\end{itemize}

The one-loop effective action $W$ is 
given by the path integral of the second order action $\mathcal{S}_2$ 
with respect to fluctuations around the background as 
\[
 W \;=\; - \ln \int\! [dY][dA][d\bar{C}][dC][d\psi]
\exp\left(- \mathcal{S}_2\right)\,. 
\]
The action $\mathcal{S}_2$ has been already diagonalized and so 
we can easily integrate out the massive degrees of freedom. 
The contribution of the fluctuations $Y$'s is given by 
\begin{eqnarray}
 W_{Y} &=&  - \ln
\Biggl[
{\rm Det} \left[ - \partial_{\tau}^2 + \frac{\mu^2}{9} \right]^{-3/2}
{\rm Det} \left[- \partial_{\tau}^2 + \frac{\mu^2}{9} \right]^{-1/2}
{\rm Det} \left[- \partial_{\tau}^2 + \frac{2}{9}\mu^2 \right]^{-3/2} 
\nn \\
&& 
\times {\rm Det} \left[- \partial_{\tau}^2 + \frac{4}{9}\mu^2 \right]^{-5/2} 
{\rm Det} \left[- \partial_{\tau}^2 + \frac{\mu^2}{36} \right]^{-6/2}
{\rm Det} \left[- \partial_{\tau}^2 + \frac{\mu^2}{4} \right]^{-18/2}
\Biggr]\,.
\end{eqnarray}
The contributions of the gauge field $A$ and the ghost $C$ $(\bar{C})$ are 
respectively given by 
\begin{eqnarray}
 W_{A} &=& - \ln
{\rm Det} \left[- \partial_{\tau}^2 + \frac{2}{9}\mu^2 \right]^{-3/2}\,, \\
 W_{\rm gh} &=& - \ln
{\rm Det} \left[- \partial_{\tau}^2 + \frac{2}{9}\mu^2 \right]^{3}\,.
\end{eqnarray}
Finally, we shall consider fermion parts. When we note that 
the $\varphi_i$'s are real, the contributions of fermions are written by
using the Pfaffian as follows:   
\begin{eqnarray}
 W_{\rm F} &=& - \ln \Biggl[ {\rm Pf} \left[
i\partial_{\tau} + \frac{\mu}{4}\gamma_{\s 123}
\right]
{\rm Pf} \left[
i\partial_{\tau} + \frac{7}{12}\mu\gamma_{\s 123}
\right]^2
{\rm Pf} \left[
i\partial_{\tau} - \frac{5}{12}\mu\gamma_{\s 123}
\right]
\Biggr]
\,,
\end{eqnarray}
where $[{\rm Pf}(B)]^2 = {\rm Det}(B)$ for an arbitrary matrix $B$. 
Here, we use the following identity, 
\[
 \ln{\rm Det}(\cdots) \;=\; {\rm Tr}'{\rm Ln}\det(\cdots)\,,
\]
where ${\rm Det}$, ${\rm Ln}$ and ${\rm Tr}'$ are the 
functional determinant, logarithm and trace, respectively. 
By the use of this identity, 
we can rewrite the expression $\ln{\rm Det}[-\partial_{\tau}^2 + M^2]$ 
as 
\begin{eqnarray}
 \ln{\rm Det}[-\partial_{\tau}^2 + M^2] \;=\;
L \int\! \frac{dk}{2\pi}\, \ln (k^2 + M^2) \;=\; L(|M| + E_{\infty})\,,
\end{eqnarray}
where $L$ is the length of the temporal direction and 
$E_{\infty}$ is a divergent constant defined by 
\[
 E_{\infty} \;\equiv\; \int\!\frac{dk}{2\pi}\ln k^2\,.
\]
Thus, we can easily obtain the one-loop effective potential $V$ by 
dividing the one-loop effective action $W$ by the length $L$ 
as $V = W/L$\,. 
The contribution of the fluctuations $Y$'s to the effective potential 
is calculated as
\begin{eqnarray}
V_{Y} &=&  \left(\frac{22}{3} + \frac{\sqrt{2}}{2} 
\right)\mu + 18 E_{\infty}\,.
\end{eqnarray}
The contribution of $A$'s is evaluated as
\begin{eqnarray}
V_{A} &=& \frac{\sqrt{2}}{2}\mu + \frac{3}{2}E_{\infty}\,,
\end{eqnarray}
and that of ghosts is expressed as
\begin{eqnarray}
V_{\rm gh} &=& - \sqrt{2}\mu - 3 E_{\infty}\,.
\end{eqnarray}
In the same way, the fermion part can be rewritten as 
\begin{eqnarray}
-\ln {\rm Pf}\left[
i\partial_{\tau} + M\gamma_{\s 123}
\right]                 
&=& - \frac{1}{2}\ln{\rm Det}\left[
i\partial_{\tau} + M\gamma_{\s 123}
\right]  \nn \\
&=& - \frac{1}{2}L\int\! \frac{dk}{2\pi}\, \ln\det 
\left(k + M\gamma_{\s 123}\right) \nn \\
&=& - 
\frac{1}{4}L\int\!\frac{dk}{2\pi}\,\left(
\ln\det\left[k + M\gamma_{\s 123}\right]
+ \ln\det\left[k - M\gamma_{\s 123}\right]
\right) \nn \\
&=& - \frac{1}{4}L\int\!\frac{dk}{2\pi}\,\ln\det\left[(
k^2 + M^2)\,{\bf 1}_{16}
\right] \nn \\
&=& - 4 L \int\!\frac{dk}{2\pi}\,\ln\left[
k^2 + M^2\right] \;=\; - 4 L (|M| + E_{\infty}) \,. \nn 
\end{eqnarray}
Thus, the one-loop contribution from the fermions is  obtained as
\begin{eqnarray}
 V_{\rm F} &=& - \frac{22}{3}\mu - 16E_{\infty}\,.
\end{eqnarray}

Therefore, the net one-loop contribution is exactly canceled 
as expected from the supersymmetry when we take 
the contributions of  massless 
bosons and ghosts into account.    

Finally, we should remark that the cancellation of the one-loop
contributions is non-trivial. 
The quantum supersymmetric vacua is non-trivial in the light-cone
formulation since the supercharges do not commute with the
Hamiltonian. In particular, the zero-point energy depends on the 
definition of the vacua. The one-loop contribution in the BMN matrix
model is essentially the zero-point energy induced 
by the pp-wave background. From such a reason 
the cancellation of one-loop contributions would be also non-trivial.

\subsection{Instability of Fuzzy Sphere}

Here, let us discuss the case of a non-supersymmetric 
fuzzy sphere with finite energy. In this case the background is 
described by  
\begin{equation}
 B^{\s I} \;=\; \frac{\mu}{6}J^{\s I}\,,\quad B^{\s I'} \;=\; 0\,.
\end{equation}
Then, the second order action $\mathcal{S}_2$ is written as 
\begin{eqnarray}
\mathcal{S}_2 &=& \int\!d\tau\, \Biggl[     
\frac{1}{2}(\dot{Y}_0^{\s I})^2 
+ \frac{1}{2}\left(\frac{\mu^2}{9}\right)(Y_0^{\s I})^2 
+ \frac{1}{2}\left((\dot{Y}_1^{\s I})^2 
+ (\dot{Y}_2^{\s I})^2 
+ (\dot{Y}_3^{\s I})^2\right) \nn \\
&& 
- \frac{\mu^2}{9}
(Y_1^1Y_2^2 + Y_2^2Y_3^3 + Y_3^3Y_1^1 - Y_1^2Y_2^1 - Y_3^2Y_2^3 
- Y_1^3Y_3^1) 
\nn \\
&&
+ \frac{1}{2}\left(\frac{\mu^2}{6}\right)
\left((Y_1^{\s I})^2 + (Y_2^{\s I})^2 + (Y_3^{\s I})^2 \right) 
+ \frac{1}{2}(\dot{Y}_0^{\s I'})^2 
+ \frac{1}{2}\left(\frac{\mu^2}{36}\right) (Y_0^{\s I'})^2 
\nn \\
&&
+ \frac{1}{2}\left((\dot{Y}_1^{\s I'})^2 
+ (\dot{Y}_2^{\s I'})^2 
+ (\dot{Y}_3^{\s I'})^2\right)
+ \frac{1}{2}\left(\frac{\mu^2}{12}
\right)\left((Y_1^{\s I'})^2 + (Y_2^{\s I'})^2 + (Y_3^{\s I'})^2 
\right) \nn \\ 
&& + \frac{1}{2}\left((\dot{A_0})^2 + 
(\dot{A}_1)^2 + (\dot{A}_2)^2 
+ (\dot{A}_3)^2 \right) 
+ \frac{1}{2}\left(\frac{\mu^2}{18}\right)\left(
(A_1)^2 + (A_2)^2 + (A_3)^2 
\right) \nn \\
&& + \frac{1}{2}\left(\dot{\bar{C}}_0 \dot{C}_0 
+ \dot{\bar{C}}_1\dot{C}_1 + \dot{\bar{C}}_2\dot{C}_2 
+ \dot{\bar{C}}_3\dot{C}_3\right) + \frac{1}{2}\left(\frac{\mu^2}{18}\right)
\left(\bar{C}_1C_1 + \bar{C}_2C_2 + \bar{C}_3C_3
\right) \nn \\ 
&& + i\left(\psi_0\dot{\psi}_0 + \psi_1\dot{\psi}_1 
+  \psi_2\dot{\psi}_2 + \psi_3\dot{\psi}_3 \right) 
+ \frac{\mu}{4}\left(\psi^{\s T}_0\gamma_{\s 123}\psi_0 
+ \psi^{\s T}_1\gamma_{\s 123}\psi_1 
+ \psi^{\s T}_2\gamma_{\s 123}\psi_2
+ \psi^{\s T}_3\gamma_{\s 123}\psi_3 \right) \nn \\
&& - \frac{\mu}{6}\left(
\psi_1^{\s T}\gamma_2\psi_3 
- \psi_1^{\s T}\gamma_3\psi_2 
+ \psi_2^{\s T}\gamma_3\psi_1 
- \psi_2^{\s T}\gamma_1\psi_3 
+ \psi_3^{\s T}\gamma_1\psi_2 
- \psi_3^{\s T}\gamma_2\psi_1 
\right)
\Biggr]\,.
\end{eqnarray}
In the same way as the case of the giant graviton, 
we must diagonalize non-diagonal 
nine components $Y_i^{\s I}$'s,\,$(i,I = 1,\,2,\,3)$ and fermions
$\psi_1,\,\psi_2$ and $\psi_3$ in order to integrate out them. 
After the naive diagonalization of the $9 \times 9$ matrix, 
the resulting action is obtained as
\begin{eqnarray}
 \mathcal{S}_{Z} &=& \int\!d\tau\, 
\Biggl[
\frac{1}{2}\left(\dot{Z}^1\right)^2 + \frac{1}{2}\left(- \frac{\mu^2}{18}
\right) (Z^1)^2 \nn \\
&& + \sum_{i=2}^4\left[\frac{1}{2}(\dot{Z}^i)^2 
+ \frac{1}{2}\left(\frac{\mu^2}{18} \right) (Z^i)^2
\right] 
+ \sum_{i=5}^9 \left[\frac{1}{2}(\dot{Z}^i)^2 
+ \frac{1}{2}\left(\frac{5}{18}\mu^2 \right) (Z^i)^2
\right] 
\Biggr]\,.
\end{eqnarray}
Then one tachyonic boson appears. This boson implies an unbounded 
direction in the quantum bosonic potential and this background is unstable.
Also the non-diagonal fermionic part can be rewritten as 
\begin{eqnarray}
 S_{\varphi} &=& \int\!d\tau\, \Biggl[
 i\varphi_0^{\s T}\dot{\varphi}_0 
+ i\varphi_1^{\s T}\dot{\varphi}_1 
+ i\varphi_2^{\s T}\dot{\varphi}_2 
+ i\varphi_3^{\s T}\dot{\varphi}_3 \nn \\ 
&& + \frac{\mu}{4}\varphi_0^{\s T}\gamma_{\s 123}\varphi_0
+ \frac{5}{12}\mu\varphi_1^{\s T}\gamma_{\s 123}\varphi_1 
+ \frac{5}{12}\mu\varphi_2^{\s T}\gamma_{\s 123}\varphi_2
- \frac{\mu}{12}\varphi_3^{\s T}\gamma_{\s 123}\varphi_3
\Biggr]\,.
\end{eqnarray}
Finally, we can obtain the following contents of fields:
\vspace*{0.5cm}
\\
\underline{Bosonic Fluctuation $Y^r$}
\begin{itemize}
\item 3 massive bosons with mass $\displaystyle \frac{\mu}{3}$:\quad
      $Y_0^{\s I}$\,,  
\item 1 tachyonic bosons with imaginary mass, whose absolute value is 
 $\displaystyle \frac{\sqrt{2}}{6}\mu$:
\quad $Z^1$\,,
\item 3 massive bosons with mass 
$\displaystyle \frac{\sqrt{2}}{6}\mu$:
\quad $Z^i\quad (i=2,\,3,\,4)$\,, 
\item 5 massive bosons with mass $\displaystyle \frac{\sqrt{10}}{6}\mu$:
\quad  $Z^i\quad (i=5,\,\ldots,\,9)$\,,
\item 6 massive bosons with mass $\displaystyle \frac{\mu}{6}$:
\quad $Y_0^{\s I'}$\,, 
\item 18 massive bosons with mass $\displaystyle \frac{\sqrt{3}}{6}\mu$:
\quad $Y_{i}^{\s I'}\quad 
(i=1,\,2,\,3)$\,,  
\end{itemize}
\underline{Gauge Field $A$}
\begin{itemize}
\item 1 massless boson:\quad $A_0$\,,
\item 3 massive bosons with mass $\displaystyle \frac{\sqrt{2}}{6}\mu$:
\quad $A_{i} \quad (i=1,\,2,\,3) $\,,
\end{itemize}
\underline{Ghost $C$}
\begin{itemize}
\item 1 massless complex ghost:\quad $C_0$\,,
\item 3 massive complex 
ghosts with mass $\displaystyle \frac{\sqrt{2}}{6}\mu$:
\quad $C_i \quad (i=1,\,2,\,3)$\,,
\end{itemize}
\underline{Fermion $\varphi$}
\begin{itemize}
\item 16 massive fermions with mass $\displaystyle \frac{\mu}{4}$:
\quad $\varphi_0$\,, 
\item 32 massive fermions with mass $\displaystyle \frac{5}{12}\mu$:
\quad $\varphi_1$ and $\varphi_2$\,,
\item 16 massive fermions with mass $\displaystyle -\frac{\mu}{12}$:
\quad $\varphi_3$\,.
\end{itemize}
Thus, we can integrate out these fields, 
and one-loop contributions to the effective potential from $Y$'s, $A$'s,
$C$'s and $\varphi$'s are respectively given by
\begin{eqnarray}
 V_{Y} &=& \left[1 + \frac{\sqrt{2}}{4} + \frac{3}{2}\sqrt{3} 
+ \frac{5}{12}\sqrt{10} + i \frac{\sqrt{2}}{12}\right]\mu + 
18 E_{\infty}\,,
\end{eqnarray}
\begin{eqnarray}
 V_{A} &=& \frac{\sqrt{2}}{4} + \frac{3}{2}E_{\infty}\,,
\end{eqnarray}
\begin{eqnarray}
 V_{\rm gh} &=& - \frac{\sqrt{2}}{2} - 3E_{\infty}\,,
\end{eqnarray}
\begin{eqnarray}
 V_{\rm F} &=& - \frac{14}{3} - 16E_{\infty}\,,
\end{eqnarray}
where we have used the formula 
\[
 \int\!\frac{dk}{2\pi}\ln(k^2 - M^2) \;=\; 
i |M| + E_{\infty}\,.
\]
The divergent constant $E_{\infty}$ is certainly
canceled by taking the massless degrees of freedom into account. 
Thus, the net contribution 
is complex and is given by 
\begin{eqnarray}
 V_{\rm net} &=& \left[
- \frac{11}{3} + \frac{3}{2}\sqrt{3} 
+ \frac{5}{12}\sqrt{10} + i \frac{\sqrt{2}}{12}\right]\mu\,.
\end{eqnarray}
The imaginary part in the effective action implies 
the instability of the non-supersymmetric fuzzy sphere background  
coming from the fact that the potential of bosons has an unbounded 
direction. 
In conclusion a non-supersymmetric fuzzy sphere background is unstable 
at the one-loop level.

\section{Conclusions and Discussions}

In this paper we have discussed classical solutions 
in the BMN matrix model. The classical solutions concretely considered 
here 
are a supersymmetric fuzzy sphere (giant graviton), 
a non-supersymmetric fuzzy sphere and oscillating gravitons. 
In particular, we have calculated the one-loop effective action 
around the supersymmetric fuzzy sphere (giant graviton) 
background and explicitly proven the quantum stability of the giant graviton 
at the one-loop level. The one-loop contributions are distinctly
canceled as expected from the requirement of supersymmetries. 
However, its cancellation is non-trivial since 
the quantum supersymmetric vacua is non-trivial in the light-cone
formulation. It is based on the fact that 
the supercharges do not commute with the
Hamiltonian. In particular, the zero-point energy depends on the 
definition of the vacua and the one-loop contribution in the BMN matrix
model is essentially the zero-point 
energy induced by the pp-wave background. Thus
the cancellation of one-loop contributions 
would be also non-trivial.     
In addition, we have discussed the instability of the
non-supersymmetric fuzzy sphere background. In this case 
the one-loop contribution deserves not to be canceled but also 
a tachyonic boson appears. 
This result implies that the quantum bosonic potential has an unbounded 
direction, and so this non-supersymmetric background is unstable 
at the one-loop level. 
In this paper, we have considered the $N=2$ case only. The extension 
to an arbitrary $N$ is an interesting future work.  

Moreover, we have discussed oscillating graviton solutions. 
Supergravitons in the flat space corresponds to 
 oscillating modes on the pp-wave 
due to the presence of mass terms. Certainly, oscillating gravitons 
turn into the usual supergravitons in the flat limit $\mu\rightarrow 0$. 
In particular, oscillating gravitons
are non-supersymmetric 
and non-BPS states, in contrast with the supergravitons in the flat space. 
Further considerations about these oscillating gravitons, such as 
scattering amplitudes 
are most interesting problems.

Other classical solutions have been already known \cite{bak}. 
In particular, the quantum stability of 
rotating 1/4 BPS solutions is so interesting but it is difficult 
to treat such a problem in the background field method 
because of their time dependences. 
It is also an interesting open problem.

\vspace{0.5cm}
\noindent 
{\bf\large Acknowledgement}

The work of K.S. is supported in part by the Grant-in-Aid from the 
Ministry of Education, Science, Sports and Culture of Japan 
($\sharp$ 14740115). 

\newpage 

\vspace*{0.5cm}
\noindent
{\bf\large Appendix}

\appendix

\section{Determinant of Bosons}

We have to calculate determinants of 9 non-diagonal bosons 
($Y_1^1$,\, $Y_1^2$,\, $Y_1^3$,\, $Y_2^1$,\,$Y_2^2$,\, $Y_2^3$,\,$Y_3^1$,\, 
$Y_3^2$, and $Y_3^3$). 
The expression of associated parts for a supersymmetric 
fuzzy sphere background is given by 
\begin{eqnarray}
 \mathcal{S}_2 &=& \int\!d\tau\,\Biggl[
\frac{1}{2}(\dot{Y}_1^{\s I})^2 
+ \frac{1}{2}(\dot{Y}_2^{\s I})^2 
+ \frac{1}{2}(\dot{Y}_3^{\s I})^2 
+ \frac{1}{2}\left(\frac{\mu^2}{3}\right)
\left((Y_1^{\s I})^2 + (Y_2^{\s I})^2 + (Y_3^{\s I})^2\right)
\nn \\
&& 
- 2\cdot\frac{\mu^2}{18}
(Y_1^1Y_2^2 + Y_2^2Y_3^3 + Y_3^3Y_1^1 - Y_1^2Y_2^1 - Y_3^2Y_2^3 
- Y_1^3Y_3^1)\Biggr]\,,
\nn 
\end{eqnarray}
and that for a non-supersymmetric background is expressed as
\begin{eqnarray}
 \mathcal{S}_2 &=& \int\!d\tau\,\Biggl[
\frac{1}{2}(\dot{Y}_1^{\s I})^2 
+ \frac{1}{2}(\dot{Y}_2^{\s I})^2 
+ \frac{1}{2}(\dot{Y}_3^{\s I})^2 
+ \frac{1}{2}\left(\frac{\mu^2}{6}\right)
\left((Y_1^{\s I})^2 + (Y_2^{\s I})^2 + (Y_3^{\s I})^2\right)
\nn \\
&& 
- \frac{\mu^2}{9}
(Y_1^1Y_2^2 + Y_2^2Y_3^3 + Y_3^3Y_1^1 - Y_1^2Y_2^1 - Y_3^2Y_2^3 
- Y_1^3Y_3^1)\Biggr]\,.
\nn 
\end{eqnarray}
We shall set $\displaystyle a = -\frac{1}{2}\partial_{\tau}^2 +
\frac{\mu^2}{6}$ (for the supersymmetric fuzzy sphere) 
or $\displaystyle 
a = -\frac{1}{2}\partial_{\tau}^2 
+ \frac{\mu^2}{12} $ (for the non-supersymmetric fuzzy sphere), 
and $\displaystyle b = \frac{\mu^2}{18}$ (for both cases).  
Then we can easily calculate the associated 
determinant by the use of the Mathematica, 
and the result is obtained as
\[
 {\rm Det} \left[(a - 2b) \cdot (a -b)^3 \cdot (a + b)^5 \right] \,.
\]

\section{Determinant of Fermions}

We shall express the fermionic parts in terms of a matrix representation as 
\begin{eqnarray}
\begin{pmatrix}
 \psi_0^{\s T}, & \psi_1^{\s T}, & \psi_2^{\s T}, & \psi_3^{\s T}
\end{pmatrix}
\begin{pmatrix}
\,a & 0 & 0 & 0 \\
\,0 & a & b\gamma_3 & -b \gamma_2 \\
\,0 & -b \gamma_3 & a & b\gamma_1 \\
\,0 & b\gamma_2 & -b\gamma_1 & a 
\end{pmatrix}
\begin{pmatrix}
\psi_0 \\
\psi_1 \\
\psi_2 \\
\psi_3 
\end{pmatrix}\,, \nn 
\end{eqnarray}
where $\displaystyle a = i\partial_{\tau} + \frac{\mu}{4}\gamma_{\s 123}$, 
and $\displaystyle 
b = \frac{\mu}{3}$ for a supersymmetric fuzzy sphere and 
$\displaystyle b = \frac{\mu}{6}$ for a non-supersymmetric fuzzy sphere.

Let us decompose the above matrix into block parts  
\begin{eqnarray}
M &\equiv&
\left(
\begin{array}{@{\,}cc|cc@{}}
a & 0 & 0 & 0 \\
0 & a & b\gamma_3 & -b\gamma_2 \\ \hline
0 & -b \gamma_3 & a & b\gamma_1 \\
0 & b\gamma_2 & -b\gamma_1 & a 
\end{array} 
\right) \;\equiv\; 
\begin{pmatrix}
A & C \\
D & B 
\end{pmatrix} 
\,. \nn 
\end{eqnarray}
By the use of the standard formula 
\begin{equation}
 {\rm Det} M = {\rm Det} A \cdot {\rm Det}(B - DA^{-1}C)\,, \label{formula}
\end{equation}
${\rm Det} M$ is rewritten as 
\begin{eqnarray}
 {\rm Det} M &=& {\rm Det} 
\left(
\begin{array}{@{\,}c|c@{}}
a^2 + b^2 & ba\gamma_1 + b^2\gamma_2\gamma_3  \\ \hline
-(ba\gamma_1 + b^2\gamma_2\gamma_3) & a^2 + b^2  
\end{array} 
\right)\,. \nn 
\end{eqnarray}
By the use of the formula (\ref{formula}) once again, we obtain the 
final result as follows:   
\begin{eqnarray}
 {\rm Det} M &=& {\rm Det}\left[(a^2 + b^2)^2 + (ba\gamma_1 +
 b^2\gamma_2\gamma_3)^2
\right] \nn \\ 
&=& {\rm Det}\left[a \cdot (a + b\gamma_{\s 123})^2 
\cdot (a - 2b\gamma_{\s 123})\right]\,. \nn 
\end{eqnarray}

\end{document}